\documentclass{book}

\usepackage[paperwidth=17cm, paperheight=24cm, includeheadfoot,
            left=2cm, right=2cm, top=2cm, bottom=1.3cm, twoside]{geometry}

\usepackage%
 {
  algorithm2e, amsmath, amssymb, appendix, caption, cite, enumitem, fancyhdr,
  fancyvrb, graphicx, import, lineno, longtable, remreset, shorttoc, sidecap,
  url
 }

\newcommand{\tmpstring}{}
\newcommand{\settmpstring}[1]{\renewcommand{\tmpstring}{#1}}
\newcommand{\SourceCodeLines}[1]
 {%
  \settmpstring{{\ttfamily\bfseries\tiny\theFancyVerbLine}}
  \ifnum#1>9
    \settmpstring
     {\parbox[b]{7.5pt}{\ttfamily\bfseries\tiny\rightline\theFancyVerbLine}}
  \fi
  \ifnum#1>99
    \settmpstring
     {\parbox[b]{11.2pt}{\ttfamily\bfseries\tiny\rightline\theFancyVerbLine}}
  \fi
 }

\def\thepart{\Alph{part}}

\makeatletter
\@removefromreset{figure}{chapter}
\makeatother

\makeatletter
\renewcommand{\thefigure}{\@arabic\c@figure}
\makeatother

\makeatletter
\@removefromreset{table}{chapter}
\makeatother

\makeatletter
\renewcommand{\thetable}{\@arabic\c@table}
\makeatother

\makeatletter
\@removefromreset{equation}{chapter}
\makeatother

\makeatletter
\renewcommand{\theequation}{\@arabic\c@equation}
\makeatother

\usepackage{tocloft}
\usepackage{minitoc}

\cftsetpnumwidth{6.1mm}

\makeatletter
\renewcommand{\@tocrmarg}{4em}
\makeatother

\mtcsetformat{minitoc}{tocrightmargin}{2.55em plus 1fil}

\newcommand{\Author}{}
\newcommand{\AuthorLastName}[1]{\renewcommand{\Author}{#1}}

\def\Title#1{\chapter[\thepart\thelecture\ \protect\mbox{\protect%
\parbox[t]{110mm}{#1 \textit{(\Author)}}}\smallskip]{\Large #1}}

\newsavebox{\shortTitleBox}
\def\shortTitle#1{\savebox{\shortTitleBox}{#1 \textit{(\Author)}}}

\newcounter{lecture}[part]

\fancypagestyle{plain}
 {
  
  \fancyhead[OL]
   {
    \vspace{-8.5pt}
    \textit{Lecture given at the International Summer School
   \emph{Modern Computational Science}}\\
    \textit{(August 9-20, 2010, Oldenburg, Germany)}
  }
 }

 {
  
  \fancyhf{}
  \fancyhead[OR]{\rightmark}
  \fancyfoot[OR]{\thepage}
  \fancyfoot[EL]{\thepage}
  \fancyhead[EL]{\usebox{\shortTitleBox}}
 }

\makeatletter
\def\thebibliography#1
{
 \section*{References\@mkboth{References}{References}}
 \list{[\arabic{enumi}]}
  {
   \settowidth\labelwidth{[#1]}\leftmargin\labelwidth\advance\leftmargin
   \labelsep\usecounter{enumi}
  }
 \def\newblock{\hskip .11em plus .33em minus .07em}
 \sloppy\clubpenalty4000\widowpenalty4000\sfcode`\.=1000\relax
}
\makeatother

\makeatletter
\renewcommand{\@makefntext}[1]{\setlength{\parindent}{0pt}%
\begin{list}{}{\setlength{\labelwidth}{1.5em}%
\setlength{\leftmargin}{\labelwidth}%
\setlength{\labelsep}{3pt}\setlength{\itemsep}{0pt }%
\setlength{\parsep}{0pt}\setlength{\topsep}{0pt}%
\footnotesize}\item[\hfill\@makefnmark]#1%
\end{list}}
\makeatother

\begin{document}

\dominitoc

\faketableofcontents

\renewcommand{\cftsecfont}{\bfseries}
\renewcommand{\cftsecleader}{\bfseries\cftdotfill{\cftdotsep}}
\renewcommand{\cftsecpagefont}{\bfseries}

\setlength{\cftsubsecindent}{12.5mm}

\captionsetup{width=0.9\textwidth,font=small,labelfont=bf}


\stepcounter{lecture}
\setcounter{figure}{0}
\setcounter{equation}{0}
\setcounter{table}{0}


\AuthorLastName{Katzgraber}

\Title{Random Numbers in Scientific Computing:\\ An Introduction}

\shortTitle{Random Numbers}

\bigskip
\bigskip


\begin{raggedright}
  \itshape Helmut G.~Katzgraber\\
  \bigskip
  Department of Physics and Astronomy, Texas A\&M University\\
  College Station, Texas 77843-4242 USA\\
  \medskip
  Theoretische Physik, ETH Zurich\\
  CH-8093 Zurich, Switzerland\\
  \bigskip
  \bigskip
\end{raggedright}


\paragraph{Abstract.}

Random numbers play a crucial role in science and industry. Many
numerical methods require the use of random numbers, in particular
the Monte Carlo method. Therefore it is of paramount importance to
have efficient random number generators. The differences, advantages
and disadvantages of true and pseudo random number generators are
discussed with an emphasis on the intrinsic details of modern and
fast pseudo random number generators.  Furthermore, standard tests
to verify the quality of the random numbers produced by a given
generator are outlined. Finally, standard scientific libraries with
built-in generators are presented, as well as different approaches
to generate nonuniform random numbers. Potential problems that one
might encounter when using large parallel machines are discussed.


\minitoc

\section{Introduction}
\label{sec:intro}

Random numbers are of paramount importance. Not only are they needed
for gambling, they find applications in cryptography, statistical
data sampling, as well as computer simulation (e.g., Monte Carlo
simulations). In principle, they are needed in any application
where unpredictable results are required. For most applications
it is desirable to have fast random number generators (RNGs) that
produce numbers that are as random as possible.  However, these two
properties are often inversely proportional to each other: excellent
RNGs are often slow, whereas poor RNGs are typically fast.

In times where computers can easily perform $10^9$ operations per
second, vast amounts of uncorrelated random numbers have to be produced
quickly. For this purpose, pseudo random number generators (PRNGs) have
been developed. However, for ``mission-critical'' applications (e.g.,
data encryption) true random number generators (TRNGs) should be used.
Both TRNGs and PRNGs have pros and cons which are outlined below.

The goal of this tutorial is to present an overview of the different
types of random number generators, their advantages and disadvantages,
as well as how to test the quality of a generator. There will be no
rigorous proofs. For a detailed mathematical treatment, the reader is
referred to Refs.~\cite{knuth:81} and \cite{press:95}. In addition,
methods to produce random numbers beyond uniform distributions are
presented. Finally, some well-tested RNG implementations in scientific
libraries are outlined.

The list of highlighted generators is by no means complete and some
readers might find that their generator of choice is probably not even
mentioned. There are {\em many} ways to produce pseudo random numbers.
In this tutorial we mainly mention those generators that have passed
common PRNG quality benchmarks and are fast. If you find yourself
using one of the bad generators outlined here, I highly recommend
you switch to one of the good generator mentioned below.

\section{True random number generators (TRNGs)}
\label{sec:trng}

TRNGs generally use some physical process that is unpredictable,
combined with some compensation mechanism that might remove any bias
in the process.

For example, the possibly oldest TRNG is coin tossing. Assuming that
the coin is perfectly symmetric, one can expect that both head or
tail will appear 50\% of the time (on average). This means that
``random bits'' $0$ (head) and $1$ (tail) can thus be generated and
grouped into blocks that then can be used to produce, for example,
integers of a given size (for example, 32 coin tosses can be used to
produce a 32-bit integer). If, however, the coin is not symmetric,
head might occur 45\% of the time, whereas tail might occur 55\%
of the time (and if you are really unlucky, the coin will land on
the edge \ldots). In such cases, post-processing corrections must be
applied to ensure that the numbers are truly random and unbiased.
TRNGs have typically the following advantages:

\begin{itemize}

\item[$\Box$]{True random numbers are generated.}

\item[$\Box$]{There are no correlations in the sequence of numbers --
assuming proper biasing compensation is performed.}

\end{itemize}

\noindent However, the fact that we are dealing with true random numbers,
also has its disadvantages:

\begin{itemize}

\item[$\Box$]{TRNGs are generally slow and therefore only of limited
use for large-scale computer simulations that require large amounts
of random numbers.}

\item[$\Box$]{Because the numbers are truly random, debugging of a
program can be difficult. PRNGs on the other hand can produce the
exact same sequence of numbers if needed, thus facilitating debugging.}

\end{itemize}

\noindent TRNGs are generally used for cryptographic applications,
seeding of large-scale simulations, as well as any application that
needs few but true random numbers. Selected implementations:

\begin{itemize}

\item[$\Box$]{Early approaches: coin flipping, rolling of dice,
roulette.  These, however, are pretty slow for most applications.}

\item[$\Box$]{Devices that use physical processes that are inherently
random, such as radioactive decays, thermal noise, atmospheric radio
noise, shot noise, etc.}

\item[$\Box$]{Quantum processes: idQuantique \cite{comment:idq}
produces hardware quantum random number generators using quantum optics
processes.  Photons are sent onto a semi-transparent mirror. Part of
the photons are reflected and some are transmitted in an unpredictable
way (the wonders of quantum mechanics\ldots). The transmitted/reflected
photons are subsequently detected and associated with random bits $0$
and $1$.}

\item[$\Box$]{Human game-play entropy: The behavior of human players
in massive multiplayer online (MMO) games is unpredictable. There
have been proposals to use this game entropy to generate true random
numbers.}

\item[$\Box$]{More imaginative approaches: Silicon Graphics produced
a TRNG based on a lava lamp. Called ``LavaRand,'' the hardware would
take images from the lava blobs inside a lava lamp. The randomness
is then extracted from the random shapes on the images and used to
seed a PRNG.}

\item[$\Box$]{\texttt{/dev/random/}: In Unix operating systems
\texttt{/dev/random/} is a source of randomness based on noise
collected from device drivers. Note that \texttt{/dev/random/} is
not necessarily a TRNG. However, for the Linux operating system
this is generally the case, although it has been shown that the
produced random numbers can have correlations when certain devices
are used for entropy gathering. Note that \texttt{/dev/urandom/}
is an ``unblocked'' version where the output is faster, but which
might contain less entropy, i.e., lower-quality random numbers.}

\end{itemize}

\section{Pseudo random number generators (PRNGs)}
\label{sec:prng}

PRNGs are based on algorithms. Therefore, PRNGs are deterministic and
not truly random. Advantages are:

\begin{itemize}

\item[$\Box$]{Random number generation is fast (no need for
post-processing).}

\item[$\Box$]{PRNGs do not require special hardware and therefore
are very portable.}

\item[$\Box$]{If needed, the exact sequence of seemingly random
numbers can be reproduced, e.g., for debugging purposes.}

\end{itemize}

\noindent The fact that good pseudo random numbers can be generated
quickly makes PRNGs the typical choice for scientific applications,
as well as statistical data analysis and noncritical applications
(think of Unix's \texttt{motd}).  However, the aforementioned
advantages come at a price:

\begin{itemize}

\item[$\Box$]{PRNGs have finite sequence lengths. At some point,
the numbers repeat. In large-scale simulations where many random
numbers are needed it is imperative to choose a good generator with
an astronomically large period \cite{comment:period}.}

\item[$\Box$]{The numbers produced by a PRNG can be correlated. In
particular, grouping the numbers in certain ways might produce
correlations that are otherwise invisible. Therefore, thorough tests
(discussed later) need to be performed before a PRNG is used in a
scientific application.}

\end{itemize}

\noindent The idea behind an algorithmic PRNG is to generate a sequence
of numbers $x_1$, $x_2$, $x_3$, \ldots using a recurrence of the form
\begin{equation}
x_i = f(x_{i-1}, x_{i-2}, \ldots, x_{i-n}) \,,
\label{eq:recurrence}
\end{equation}
where $n$ initial numbers (seed block) are needed to
start the recurrence.  All PRNGs have the structure shown in
Eq.~(\ref{eq:recurrence}), the magic lies in finding a function $f$
that produces numbers that are ``as random as possible.'' Some PRNGs
use the modulo operation to further randomize the numbers. This has
the effect that often the maximum sequence length is limited.

The seed determines the sequence of random numbers. {\em Therefore, it
is crucial to seed the PRNG carefully}. For example, if the period of
the PRNG is rather short, repeated seeding might produce overlapping
streams of random numbers. Furthermore, there are generators where
a poor choice of the seed might produce correlated random numbers.
So \ldots which generator should one use? In what follows, some
typical PRNGs are discussed and outlined.

\subsection{Linear congruential generators}
\label{subsec:lcg}

Linear congruential generators (LCGs) are one of the oldest PRNGs. In
their simplest implementation, they are of the form \cite{hartmann:09}
\begin{equation}
x_{i+1} = (a x_{i} + c) \bmod m
\label{eq:lcg1}
\end{equation}
with $x_0$ a seed value. In Eq.~(\ref{eq:lcg1}) $m$ is a large
integer that determines the period of the generator; it will thus
produce numbers between $0$ and $m - 1$. Note that this is similar
to a roulette where a croupier spins a wheel with $37$ pockets in one
direction, then spins a ball in the opposite direction around a tilted
circular track running around the circumference of the wheel. The ball
eventually slows down and lands in one of the $m = 37$ pockets.  $0 \le
a < m$ is called the multiplier and $0 \le c < m$ is the increment. The
case where $c = 0$ corresponds to the Park-Miller PRNG.  The values
of $a$, $c$, $x_0$ and $m$ can heavily influence the behavior of the
LCG. One can rigorously show that a LCG has period $m$ if and only
if $c$ is relatively prime to $m$, $a - 1$ is a multiple of $p$
for every prime $p$ dividing $m$, and $a - 1$ is a multiple of $4$,
if $m$ is a multiple of $4$. Dizzy yet?  An acceptable choice is
given by the GGL generator where $a = 16807$, $c = 0$ and $m = 2^{31}
- 1$. An example of a bad generator is given below.

A more general approach is used in linear feedback shift register
generators given by the following recurrence (with $c = 0$) 
\cite{mertens:09}
\begin{equation}
x_{i} = (a_1 x_{i-1} + a_2 x_{i-2} + \ldots + a_n x_{i-n}) \bmod p \,.
\end{equation}
Here $p$ is a prime number. The quality of the pseudo random numbers
depends on the multipliers $a_k$, as well as $n$ and $p$. One can show
that the maximum period of such a generator is $p^n - 1$. However,
if the parameters of the generator are not chosen carefully, the
period can be considerably shorter than the maximum period. The period
is maximal if and only if the characteristic polynomial
\begin{equation}
f(x) = x^n - a_1 x^{n-1} - a_2 x^{n-2} - \ldots - a_n
\end{equation}
is primitive modulo $p$.

LCGs are extremely fast and use little memory, however, the period
is limited by the choice of $m$. For standard LCGs, $m \sim 2^{32}$
which corresponds to approximately $10^9$ random numbers. On a modern
computer such a sequence is exhausted in seconds. If $m = 2^k$ ($k \in
{\mathbb N}$) then lower-order bits of the generated sequence have a
far shorter period than the sequence as a whole. Therefore {\em never
use a linear congruential PRNG for numerical simulations}. However,
it is acceptable to use a LCG to generate a seed block for a more
complex PRNG.  Finally, note that LCGs are difficult to parallelize.

\paragraph{Example of a bad generator: RANDU}

\begin{SCfigure}[1.2][!htb]
\includegraphics[scale=0.45]{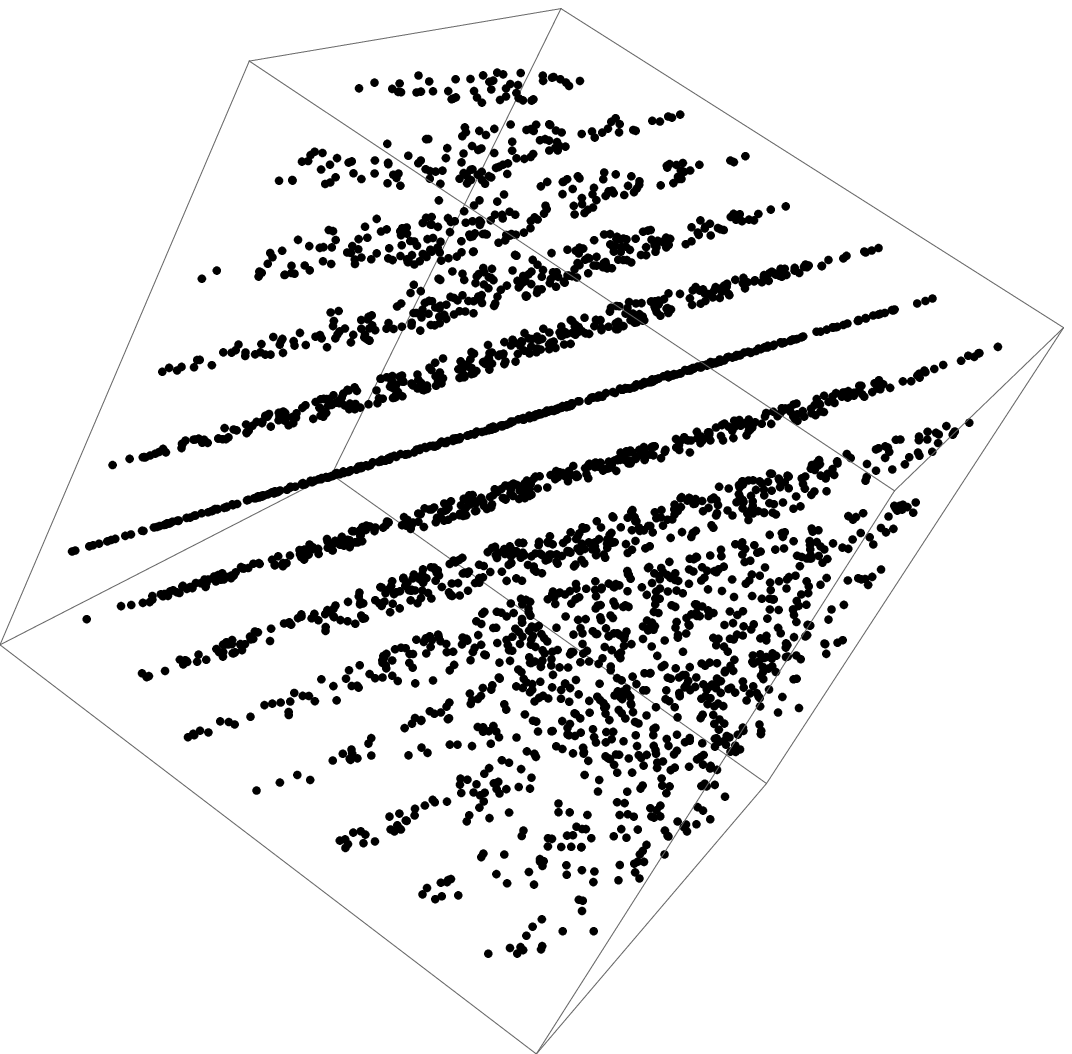}\hspace{2pc}
\caption{
$10^3$ triplets of successive random numbers produced with RANDU
plotted in three-dimensional space. If the random numbers were
perfectly random, no planes should be visible. However, when viewed
from the right angle, planes emerge, thus showing that the random
numbers are strongly correlated.
\vspace{3.0pc} }
\label{fig:randu} 
\end{SCfigure}

RANDU is a linear congruential PRNG of the Park-Miller type that was
installed as the standard generator on IBM mainframe computers in the
1960s. It uses the parameters $a = 65539$, $c = 0$, and $m = 2^{31}$.
The particular choice of $a = 65539 = 2^{16} + 3$ was made to
speed up the modulo operation on 32-bit machines. The fact that the
numbers have correlations can be illustrated with the following simple
calculation (modulo $m$ means that $2^{32} \equiv 0$)
\begin{eqnarray}
x_{i+2} &=& a x_{i+1} 
	 = (2^{16} + 3)x_{i+1} 
	 = (2^{16} + 3)^2 x_i 			\\ \nonumber
	&=& (2^{32} + 6 \cdot 2^{16} + 9) x_i 
 	 \equiv [6(2^{16} + 3) - 9]x_i		\\ \nonumber
	&=& 6x_{i+1} - 9x_i \, .
\end{eqnarray}
Therefore, tuplets of random numbers have to be correlated. If three
consecutive random numbers $x_1$, $x_2$ and $x_3$ are combined
to a vector $(x_1,x_2,x_3)$, then the numbers lie on planes in
three-dimensional space, as can be seen in Fig.~\ref{fig:randu}.

\subsection{Lagged Fibonacci generators}
\label{subsec:lfg}

Lagged Fibonacci generators are intended as an improvement over linear
congruential generators and, in general, they are not only fast but
most of them pass all standard empirical random number generator tests.
The name comes from the similarity to the Fibonacci series
\begin{equation}
x_i = x_{i - 1} + x_{i - 2} \;\;\; \longrightarrow \;\;\;
\{1,1,2,3,5,8,13,21,\ldots\}
\label{eq:fib}
\end{equation}
with $x_0 = x_1 = 1$. In this case we generalize Eq.~(\ref{eq:fib}) to the 
following sequence
\begin{equation}
x_i = (x_{i - j} \odot x_{i - k}) \bmod m\,, \;\;\;\;\;\;\;\;\;\;\; 0 < j < k,
\label{eq:fib2}
\end{equation}
where $\odot$ represents a binary operator, i.e., addition,
multiplication or exclusive OR (XOR). Typically $m = 2^M$ with $M =
32$ or $64$. Generators of this type require a {\em seed block} of
size $k$ to be initialized. In general, one uses a very good yet
possibly slow [e.g., \texttt{ran2( )}, see below] PRNG to build
the seed block. When the operator is a multiplication [addition]
the PRNG is called a multiplicative [additive] lagged Fibonacci
generator. The case where the operator is XOR is known as two-tap
generalised feedback shift register (GFSR). Note that the Mersenne
Twister (discussed below in Sec.~\ref{subsec:gens}) is a variation of
a GFSR. In fact, the linear and generalized shift register generators,
the Mersenne Twister and the WELL PRNG (see below) belong to a class
of generators known as ${\mathbb F}_2$-linear PRNGs because they
are based on a recurrence over a finite binary field ${\mathbb F}_2$.

The theory behind this class of generators is rather complex
and there are no rigorous proofs on the performance of these
generators. Therefore their quality relies vastly on statistical
tests. In particular, they are very sensitive to initialization,
which is why a very good generator has to be used to build the seed
block. Furthermore, the values of $j$ and $k$ have to be chosen
carefully. For the generator to achieve the maximum period, the
polynomial
\begin{equation}
y = x^k + x^j + 1
\end{equation}
must be primitive over the integers modulo 2. Some commonly-used
choices for 64-bit additive generators are the following pairs:
$\{55,24,\oplus\}$, $\{607,273,\oplus\}$, $\{2281,1252,\oplus\}$,
$\{9689,5502,\oplus\}$. For multiplicative generators common values
are $\{1279,418,\otimes\}$ and $\{250,103,\otimes\}$.  Note that, in
general, the larger the values of the lags, the better the generator.
Furthermore, the length of the period $\rho$ depends on $m = 2^M$
and $k$. For example:
\begin{equation}
\rho(\oplus) = 2^{k - 1}2^{M - 1}, 
\;\;\;\;\;\;\;\;\;\;\;\;\;\;\;\;\;\;\;\;\;\;\;\;\;\;\;\;\; 
\rho(\otimes) = 2^{k - 1}2^{M - 3} \, .
\end{equation}
Lagged Fibonacci generators are thus fast, generally pass all known
statistical quality tests  and have very long periods. They can
also be vectorized on vector CPU computers, as well as pipelined on
scalar CPUs.

\paragraph{Example of a commonly-used good generator: r1279} In the
case of \texttt{r1279( )} with 32 bits---a multiplicative generator
with $k = 1279$---the period is approximately $10^{394}$, a {\em
very} large number if you compare to linear congruential generators.
\texttt{r1279( )} passes all known RNG tests. Furthermore, there are
fast implementations. Therefore, it is one of the RNGs of choice in
numerical simulations, which is why it is standard in many scientific
computing libraries, such as the GSL \cite{comment:gsl}.

\paragraph{Example of a bad generator: r250} For many years
\texttt{r250( )} ($k = 250$, $\odot = {\rm XOR}$) was the standard
generator in numerical simulations. Not only was it fast, it passed
all common RNG quality tests at that time. However, in 1992 Ferrenberg
{\em et al.}~performed a Monte Carlo simulation of the two-dimensional
Ising model \cite{ising:25,huang:87} using the Wolff cluster algorithm
\cite{wolff:89}.  Surprisingly, the estimate of the energy per spin at
the critical temperature was approximately $42$ standard deviations
off the known exact result. After many tests they concluded that
the random number generator used, namely \texttt{r250( )} was the
culprit. This case illustrates that although a generator passes
all known statistical tests, there is {\em no guarantee} that the
produced numbers are random enough.

\subsection{Other commonly-used PRNGs}
\label{subsec:gens}

\paragraph{Mersenne Twister} The Mersenne Twister was developed in 1997
by Matsumoto and Nishimura and is a version of a generalised feedback
shift register PRNG. The name comes from the fact that the period is
given by a Mersenne prime ($M_n = 2^n - 1$, $n \in {\mathbb N}$). It is
very fast and produces high-quality random numbers. The implementation
\texttt{mt19937( )}, which is part of many languages and scientific
libraries such as Matlab, R, Python, Boost \cite{comment:boost} or the
GSL \cite{comment:gsl}, has a period of $\rho = 2^{19937} - 1 \approx
10^{6001}$. There are two common versions of \texttt{mt19937( )} for
32 and 64-bit architectures. For a $k$-bit word length, the Mersenne
Twister generates numbers with a uniform distribution in the range
$[0,2^k - 1]$. Although the Mersenne Twister can be checkpointed
easily, it is based on a rather complex algorithm.

\paragraph{WELL generators} The name stands for Well Equidistributed
Long-period Linear. The idea behind the generator originally
developed by Panneton, L'Ecuyer and Matsumoto is to provide better
equidistribution and bit mixing with an equivalent period length and
speed as the Mersenne Twister.

\paragraph{ran2} The Numerical Recipes \cite{press:95} offers different
random number generators. Do {\em not} use the quick-and-dirty
generators for mission critical applications. They are quick, but
dirty (and thus bad). Both \texttt{ran0( )} and \texttt{ran1( )}
are not recommended either since they do not pass all statistical
tests and have short periods of $2^{32}$. \texttt{ran2(\ \nolinebreak)},
however, has a period of $\sim 10^{18}$ (still modest in comparison to other
generators outlined here) and passes all statistical tests. In fact,
the authors of the Numerical Recipes are willing to pay \$1000 to the
first person who proves that \texttt{ran2( )} fails a statistical
test. Note that \texttt{ran2( )} is rather slow and should only be
used to generate seed blocks for better generators.

\paragraph{drand48} The Unix built-in family of generators
\texttt{drand48( )} is actually based on a linear congruential
generator with a 48-bit integer arithmetic. Pseudo random numbers are
generated according to Eq.~(\ref{eq:lcg1}) with $a = 25214903917$,
$c = 11$ and $m = 2^{48}$. Clearly, this generator should not be
used for numerical simulations. Not only is the maximal period only
$\sim 10^{14}$, linear congruential generators are known to have
correlation effects.

\paragraph{Online services} A website that delivers true random numbers
is \texttt{random.org}. Although not very useful for large-scale
simulations, the site delivers true random numbers (using atmospheric
noise). There is a limit of free random bits. In general, the service
costs approximately US\$ 1 per 4 million random bits. A large-scale
Monte Carlo simulation with $10^{12}$ random numbers would therefore
cost US\$ 250,000.

\paragraph{Final recommendation} For any scientific applications avoid
the use of linear congruential generators, the family of Unix built-in
generators \texttt{drand48( )}, Numerical Recipe's \texttt{ran0( )},
\texttt{ran1( )} and \texttt{ran2( )}, as well as any home-cooked
routines.  Instead, use either a multiplicative lagged Fibonacci
generator such as \texttt{r1279( )}, WELL generators, or the Mersenne
Twister. Not only are they good, they are very fast.

\section{Testing the quality of random number generators}
\label{sec:tests}

In the previous chapters we have talked about ``good'' and ``bad''
generators mentioning often ``statistical tests.'' There are obvious
reasons why a PRNG might be bad: For example, with a period of $10^9$
a generator is useless for most scientific applications. As in the case of
\texttt{r250( )} \cite{ferrenberg:92}, there can be {\em very} subtle effects
that might bias data in a simulation. These subtle effects can often only
be discovered by performing batteries of statistical tests that try
to find hidden correlations in the stream of random numbers. In the
end, our goal is to obtain pseudo random numbers that are like true
random numbers.

Over the years many empirical statistical tests have been developed
that attempt to determine if there are any short-time or long-time
correlations between the numbers, as well as their distribution. Are
these tests enough? No. As in the case of \texttt{r250( )}, your simulation
could depend in a subtle way on hidden correlations.
What is thus the ultimate test? {\em Run your code with different PRNGs.
If the results agree within error bars and the PRNGs used are from different
families, the results are likely to be correct}.

\subsection{Simple PRNG tests}
\label{subsec:tests1}

If your PRNG does not pass the following tests, then you should
definitely not use it. The tests are based on the fact that if one
assumes that the produced random numbers have no correlations, the
error should be purely statistical and scale as $1/\sqrt{N}$ where $N$
is the number of random numbers used for the test.

\paragraph{Simple correlations test} For all $n \in {\mathbb N}$
calculate the following function
\begin{equation}
\varepsilon(N,n) = \frac{1}{N}\sum_{i = 1}^N x_i x_{i+n} - E(x)^2 \, ,
\end{equation}
where
\begin{equation}
E(x) = \frac{1}{N}\sum_{i = 1}^N x_i
\end{equation}
represents the average over the sampled pseudo random numbers.  If the
tuplets of numbers are not correlated, $\varepsilon(N,n)$ should
converge to zero with a statistical error for $N \to \infty$, i.e.,
\begin{equation}
\varepsilon(N,n) \sim {\mathcal O}(N^{-1/2}) \;\;\;\;\;\;\;\;\; \forall n\;.
\end{equation}

\paragraph{Simple moments test} Let us assume that the PRNG to be
tested produces uniform pseudo random numbers in the interval $[0,1]$.
One can analytically show that for a uniform distribution the $k$-th
moment is given by $1/(k + 1)$. One can therefore calculate the
following function for the $k$th moment
\begin{equation}
\mu(N,k) = \left| \frac{1}{N}\sum_{i = 1}^N x_i^k - \frac{1}{k + 1}\right| \;.
\end{equation}
Again, for $N \to \infty$
\begin{equation}
\mu(N,k) \sim {\mathcal O}(N^{-1/2}) \;\;\;\;\;\;\;\;\; \forall k \; .
\end{equation}

\begin{figure}[!htb]
\centering
\includegraphics[scale=0.50]{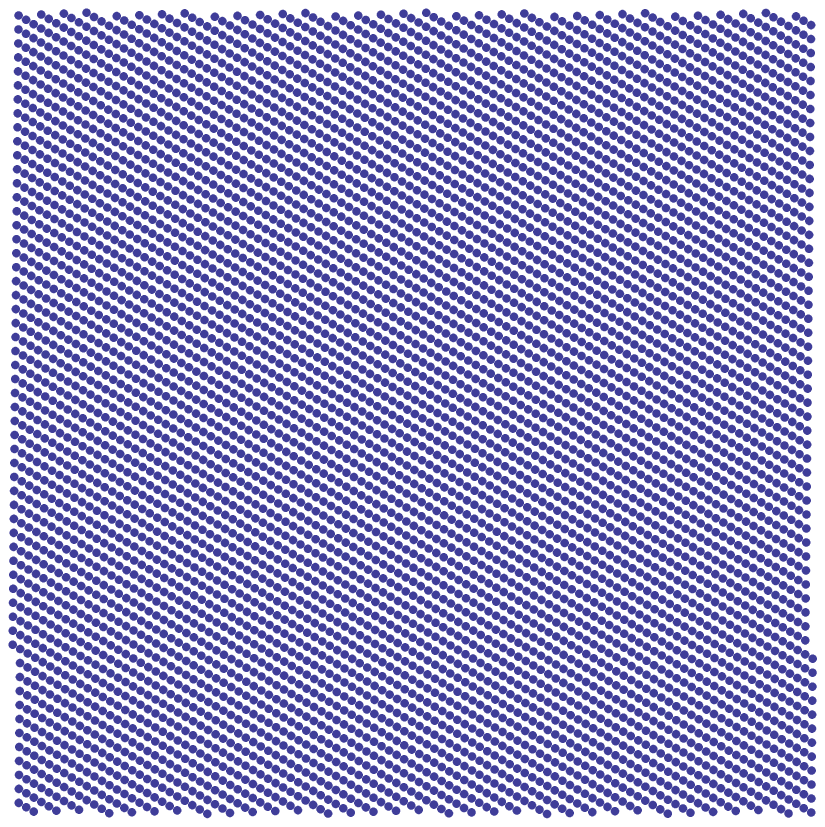}
\hspace*{1.5cm}\includegraphics[scale=0.50]{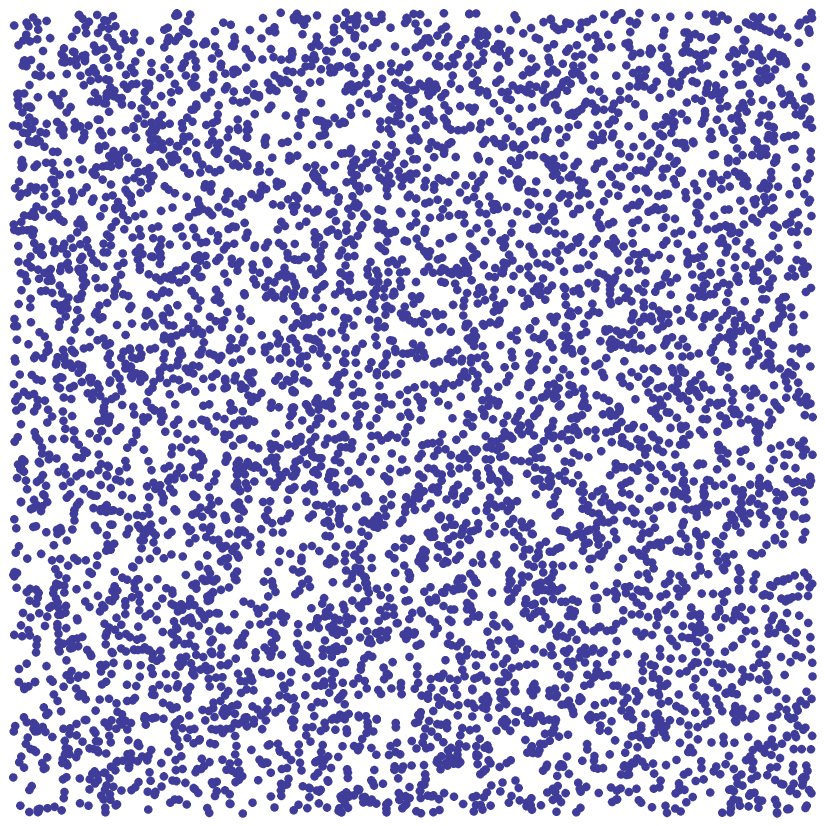}

\vspace*{1.0cm}

\includegraphics[scale=0.50]{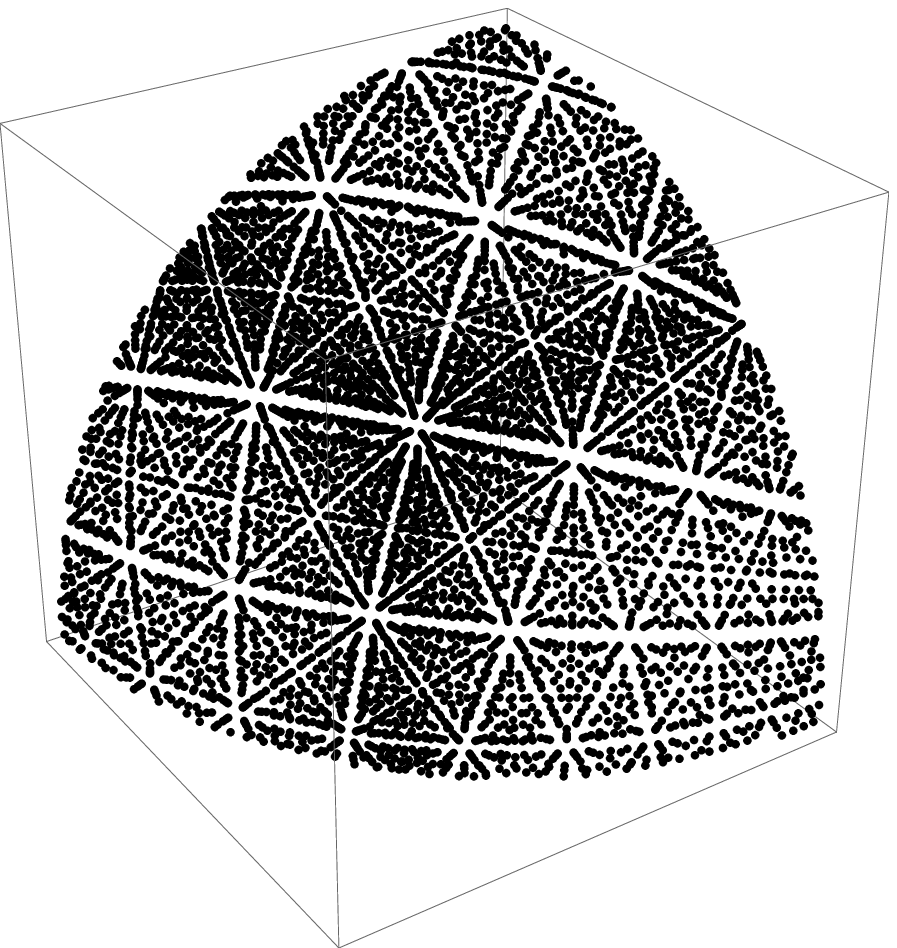}
\hspace*{1.5cm}\includegraphics[scale=0.50]{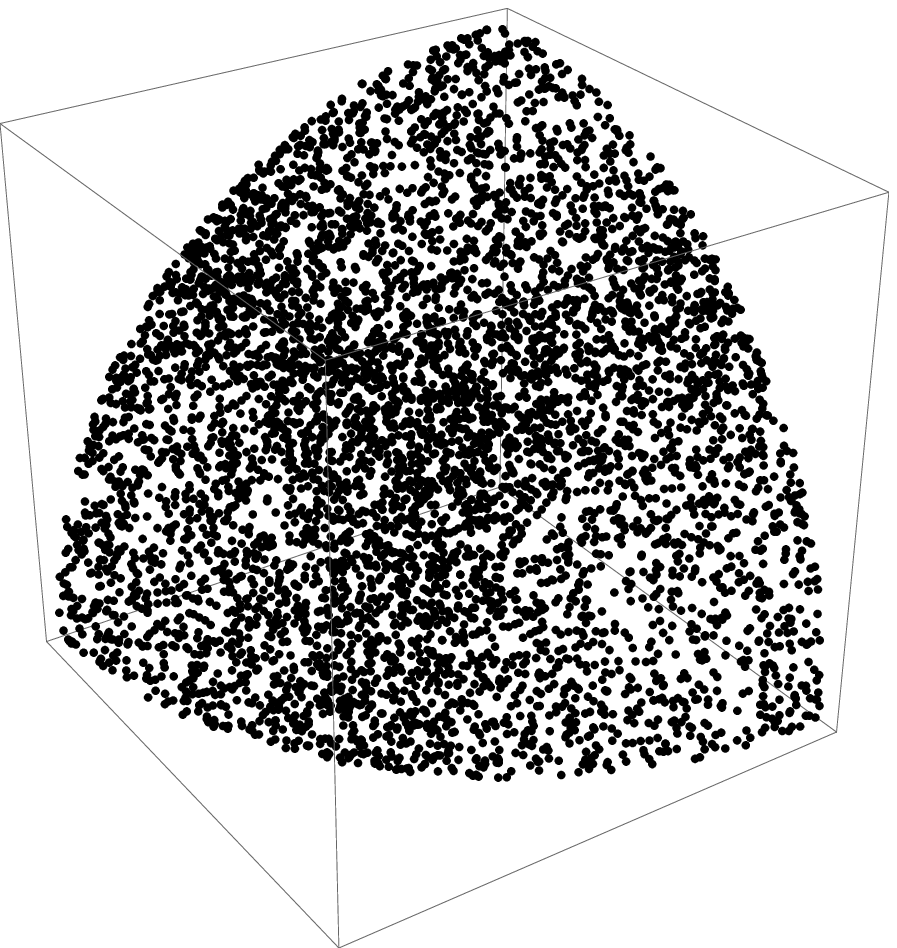}
\caption{
Graphical correlations test using a home-cooked linear congruential
generator with $a = 106$, $c = 1283$ and $m = 6075$ (left panels) and
the Mersenne Twister (right panels). The top row shows 2-tuplets in
the plane. Correlations (left-hand-side) are clearly visible for the
home-cooked PRNG. The bottom row shows 3-tuplets on the unit sphere.
Again, the home-cooked PRNG (left-hand-side), albeit pretty, shows
extreme correlations.
}
\label{fig:graphtest}
\end{figure}

\paragraph{Graphical tests} Another simple test to look for ``spatial
correlations'' is to group a stream of pseudo random numbers into
$k$-tuplets.  These tests are also known under the name ``spectral
tests.''  For example, a stream of numbers $x_1$, $x_2$, \ldots can
be used to produce two-dimensional vectors $\vec{v}_1 = (x_1,x_2)$,
$\vec{v}_2 = (x_3,x_4)$, \ldots, as well as three-dimensional or
normalized unit vectors ($\vec{e} = \vec{x}/||\vec{x}||$). Figure
\ref{fig:graphtest} shows data for 2-tuplets and normalized 3-tuplets
for both good and bad PRNGs. While the good PRNG shows no clear sign
of correlations, these are evident in the bad PRNG.

Theoretical details on spectral tests, as well as many other methods
to test the quality of PRNGs such as the chi-squared test, can be
found in Ref.~\cite{knuth:81}.

\subsection{Test suites}
\label{sec:suites}

There are different test suites that have been developed with the sole
purpose of testing PRNGs. In general, it is recommended to use these to
test a new generator as they are well established. Probably the oldest
and most commonly used test suite is DIEHARD by George Marsaglia.

\paragraph{DIEHARD} The software is freely available
\cite{comment:diehard} and comprises 16 standard tests.  Most of
the tests in DIEHARD return a $p$-value, which should be uniform
on the interval $[0,1)$ if the pseudo random numbers are truly
independent random bits. When a bit stream fails the test, $p$-values
near 0 or 1 to six or more digits are obtained (for details see
Refs.~\cite{comment:diehard} and \cite{knuth:81}). DIEHARD includes
the following tests (selection):

\begin{itemize}

\item[$\Box$]{Overlapping permutations test: Analyze sequences of
five consecutive random numbers. The $5! = 120$ possible permutations
should occur (statistically) with equal probability.}

\item[$\Box$]{Birthday spacings test: Choose random points on a large
interval. The spacings between the points should be asymptotically
Poisson-distributed.}

\item[$\Box$]{Binary rank test for $32 \times 32$ matrices: A random
$32 \times 32$ binary matrix is formed. The rank is determined, it
can be between 0 and 32. Ranks less than 29 are rare, and so their
counts are pooled with those of 29. Ranks are found for $40\,000$
random matrices and a chi-square test \cite{knuth:81} is performed
for ranks 32, 31, 30, and $\le 29$.}

\item[$\Box$]{Parking lot test: Randomly place unit circles in a $100
\times 100$ square. If the circle overlaps an existing one, choose
a new position until the circle does not overlap. After $12\,000$
attempts, the number of successfully ``parked'' circles should follow
a certain normal distribution.}

\end{itemize}

\noindent It is beyond the scope of this lecture to outline all
tests. In general, the DIEHARD tests perform operations on random
number streams that in the end should be either distributed according
to a given distribution that can be computed analytically, or the
problem is reduced to a case where a chi-square or Kolmogorov-Smirnov
test \cite{knuth:81} can be applied to measure the quality of the
random series.

\paragraph{NIST test suite} The US National Institute of Standards
and Technology (NIST) has also published a PRNG test suite
\cite{comment:nist}. It can be downloaded freely from their
website. The test suite contains 15 tests that are extremely well
documented. The software is available for many architectures and
operating systems and considerably more up-to-date than DIEHARD. Quite
a few of the tests are from the DIEHARD test suite, however, some
are novel tests that very nicely test the properties of PRNGs.

\paragraph{L'Ecuyer's test suite} Pierre L'Ecuyer has not only
developed different PRNGs, he has also designed TestU01, a ANSI C
software library of utilities for the empirical statistical testing
of PRNGs \cite{comment:testu01}. In addition, the library implements
several PRNGs and is very well documented.

\section{Nonuniform random numbers}
\label{sec:nonuniform}

Standard random number generators typically produce either bit streams,
uniform random integers between \texttt{0} and \texttt{INT\_MAX},
or floating-point numbers in the interval $[0,1)$. However, in many
applications it is desirable to have random numbers distributed
according to a probability distribution that is not uniform. In this
section different approaches to generate nonuniform random numbers
are presented.

\subsection{Binary to decimal}

Some generators merely produce streams of binary bits. Using the relation
\begin{equation}
u = \sum_{i = 0}^{B} b_i 2^i
\end{equation}
integers $x$ between $0$ and $2^B$ can be produced.  The bit stream
$b_i$ is buffered into blocks of $B$ bits and from there an integer is
constructed. If floating-point random numbers in the interval $[0,1)$
are needed, we need to replace $u \to u / 2^{B}$.

\subsection{Arbitrary uniform random numbers}

Uniform random numbers $r$ in the interval $[a,b)$ can be computed
by a simple linear transformation starting from uniformly distributed
random numbers $u \in [0,1)$
\begin{equation}
r = a + (b - a)u \, .
\end{equation}
More complex transformations need the help of probability theory.

\subsection{Transformation method}
\label{subsec:inv}

The probability $p(u)du$ of generating a uniform random number between
$u$ and $u + du$ is given by
\begin{equation}
p(u)du = \left\{ \begin{array}{rl}
 du & 0 < u < 1 \\
  0 &\mbox{ otherwise} \, .
       \end{array} \right. 
\end{equation}
Note that the probability distribution is normalized, i.e.,
\begin{equation}
\int_{-\infty}^{\infty} p(u) du = 1 \, .
\end{equation}
Suppose we take a prescribed function $y(u)$ of a uniform random number
$u$. The probability distribution of $y$, $q(y)dy$, is determined by
the transformation law of probabilities, namely \cite{hartmann:09}
\begin{equation}
|q(y)dy| = |p(u)du| \longrightarrow q(y) = p(u) \left|\frac{du}{dy}\right|\,.
\label{eq:xform}
\end{equation}
If we {\em can invert the function}, we can compute nonuniform deviates.

\subsection{Exponential deviates}

To compute exponentially-distributed random numbers with
\begin{equation}
q(y) = a\exp(-a y)
\label{eq:expd} 
\end{equation}
we use Eq.~(\ref{eq:xform}):
\begin{equation}
\left|\frac{du}{dy}\right| = a\exp(-a y) \longrightarrow u(y) = \exp(-a y) \, .
\label{eq:exp}
\end{equation}
Inverting Eq.~(\ref{eq:exp}) we obtain for exponentially-distributed random
numbers
\begin{equation}
y = -\frac{1}{a} \ln(u) \, ,
\end{equation}
where $u \to 1 - u \in (0,1]$ is a uniform random number.

\subsection{Gaussian-distributed random numbers} 
\label{subsec:gaussian}

Gaussian-distributed (also known as Normal) random numbers find
wide applicability in many computational problems. It is therefore
desirable to efficiently compute these. The probability distribution
function is given by
\begin{equation}
q(y) = \frac{1}{\sqrt{2\pi}}\exp(-y^2/2) \, .
\label{eq:gaussgauss}
\end{equation}
The most widespread approach to generate Gaussian-distributed random
numbers is the Box-Muller method: The transformation presented in
Sec.~\ref{subsec:inv} can be generalized to higher space dimensions.
In one space dimension, it is not possible to solve the integral and
therefore invert the function. However, in two space dimensions this
is possible:
\begin{equation}
q(x)q(y) dx dy = \frac{1}{2\pi}e^{-(x^2 + y^2)/2} dx dy 
= \frac{1}{2\pi}e^{-R^2/2} R dR d\theta \to e^{-t}dt \, .
\end{equation}
Let $u_1$ and $u_2$ be two uniform random numbers. Then
\begin{equation}
\theta = 2\pi u_1, \;\;\;\;\;\;\;\;\;\;\;\;\;\;\;\;\;\;\; t = -\ln(u_2)\,.
\end{equation}
It follows that
\begin{equation}
R = \sqrt{-2 \ln(u_2)}
\end{equation}
and therefore
\begin{eqnarray}
\label{eq:trig}
x &=& \sqrt{-2\ln(u_2)}\cos(2\pi u_1), \\ \nonumber
y &=& \sqrt{-2\ln(u_2)}\sin(2\pi u_1) \, .
\end{eqnarray}
At each step of the algorithm two uniform random numbers are converted
into two Gaussian-distributed random numbers. Using simple rejection
methods one can speed up the algorithm by preventing the use of
trigonometric functions in Eqs.~(\ref{eq:trig}). For details see
Ref.~\cite{press:95}.

\subsection{Acceptance-Rejection method}

When the integral in the transformation method (Sec.~\ref{subsec:inv})
cannot be inverted easily one can apply the acceptance-rejection
method, provided that the distribution function $f(x)$ for
the random numbers is known and computable. The idea behind the
acceptance-rejection method is simple: Find a distribution function
$g(x)$ that bounds $f(x)$ over a finite interval (and for which one
can easily compute random numbers):
\begin{equation}
f(x) \le g(x) \, .
\end{equation}
The algorithm is
simple \cite{hartmann:09}:

\SourceCodeLines{9}
\begin{Verbatim}[fontsize=\small]
 repeat
     generate a g-distributed random number x from g(x)
     generate a uniform random number u in [0,1]
 until
     u < f(x)/g(x)  
 done

 return x
\end{Verbatim}

\begin{SCfigure}[1.2][!htb]
\includegraphics[scale=0.50]{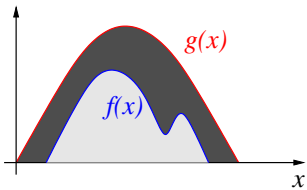}\hspace{2pc}
\caption{
Illustration of the rejection method. If the point $(u, g(x))$ lies
in the light shaded area, it is $f$-distributed. If it lies in the
dark shaded area it is rejected.
\vspace{1.0pc} }
\label{fig:reject}
\end{SCfigure}

\noindent Basically, one produces a point in the two-dimensional plane
under the function $g(x)$, see Fig.~\ref{fig:reject}. If the point lies
under the function $f(x)$ it is accepted as an $f$-distributed random
number (light shaded area in Fig.~\ref{fig:reject}). If it lies in
the dark shaded area of Fig.~\ref{fig:reject} it is rejected. Note
that this is very similar to Monte Carlo integration: The number
of rejected points depends on the ratio between the area of $g(x)$
to the area of $f(x)$. Therefore, it is imperative to have a good
guess for the function $g(x)$ that\ldots

\begin{itemize}

\item[$\Box$]{is as close as possible to $f(x)$ to prevent many
rejected moves.}

\item[$\Box$]{is quickly evaluated.}

\end{itemize}

\noindent For very complex cases numerical inversion of the function
$f(x)$ might be faster.

\subsection{Random numbers on a $N$-sphere}

Sometimes it is necessary to generate random number on a
$N$-sphere. There are two possible approaches:

\begin{itemize}

\item[$\Box$]{Using the Box-Muller method (Sec.~\ref{subsec:gaussian}):

\begin{itemize}

\item[$\Box$]{Start from a uniform random vector $\vec{u}$.}

\item[$\Box$]{Use the Box-Muller method on each component to obtain
a normally-distributed vector $\vec{n}$.}

\item[$\Box$]{Normalize the length of the vector to unity: $\vec{e}
= \vec{n}/||\vec{n}||$. The angles are now uniformly distributed.}

\end{itemize}

}

\item[$\Box$]{Using Acceptance-Rejection:

\begin{itemize}

\item[$\Box$]{Generate a uniform random vector $\vec{u}$ with each
component in the interval $[-1,1]$.}

\item[$\Box$]{If $||\vec{u}|| > 1$, choose a new vector.}

\item[$\Box$]{Otherwise normalize the length of $\vec{u} \to
\vec{u}/||\vec{u}||$.}

\end{itemize}

}

\end{itemize}

\noindent The second approach using the acceptance-rejection method
works better if $N$ is small.

\section{Library implementations of PRNGs}

It is {\em not} recommended to implement one's own PRNG, especially
because there are different well-tested libraries that contain most
of the common generators. In addition, these routines are highly
optimized, which is very important. For example, in a Monte Carlo
simulation the PRNG is the most called function (at least 80\% of
the time). Therefore it is crucial to have a fast implementation.
Standard libraries that contain PRNGs are

\begin{itemize}

\item[$\Box$]{Boost Libraries \cite{comment:boost}: Generic
implementation of many PRNGs in C++.}

\item[$\Box$]{GSL (Gnu Scientific Library) \cite{comment:gsl}:
Efficient implementation of a vast selection of PRNGs in C with
checkpointing built in.}

\item[$\Box$]{TRNG \cite{comment:trng}: Implementation of different
PRNGs with checkpointing built in. The library is designed with
large-scale parallel simulations in mind, i.e., block splitting and
leapfrogging are also implemented \cite{bauke:07,mertens:09}.}

\item[$\Box$]{Numerical Recipes \cite{press:95}: Implementation
of some PRNGs.  The libraries, however, are dated and the license
restrictions ridiculous.}

\end{itemize}

\noindent In what follows some details on how to use random number
generators on both the GSL and Boost libraries are outlined.  Note that
these libraries are built in a modular way. They contain:

\begin{itemize}

\item[$\Box$]{(Uniform) Pseudo random number generator engines (e.g.,
Mersenne Twister, LCGs, Lagged Fibonacci generators, \ldots).}

\item[$\Box$]{Distribution functions (e.g., Gaussian, Gamma, Poisson,
Binomial, \ldots).}

\item[$\Box$]{Tests.}

\end{itemize}

\subsection{Boost}

The C++ Boost libraries \cite{comment:boost} contain several PRNGs
outlined in these lecture notes. For example, one can define a PRNG
\texttt{rng1} that produces random numbers using the Mersenne Twister,
a \texttt{rng2} that produces random numbers using a lagged Fibonacci
generator, or \texttt{rng3} using a LCG with the following lines
of code:

\begin{verbatim}
 boost::mt19937 rng1;                   // mersenne twister
 boost::lagged_fibonacci1279 rng2;      // lagged fibonacci r1279
 boost::minstd_rand0 rng3;              // linear congruential
\end{verbatim}

\noindent These can now be combined with different distribution
functions. The uniform distributions in an interval $[a,b]$ can be
called with

\begin{verbatim}
 boost::uniform_int<int> dist1(a,b);     // integers between a and b
 boost::uniform_real<double> dist2(a,b); // doubles between a and b
\end{verbatim} 

\noindent There are many more distribution functions and the reader is
referred to the documentation \cite{comment:boost}. For example

\begin{verbatim}
 boost::exponential_distribution<double> dist3(a);
\end{verbatim}

\noindent produces random numbers with the distribution
shown in Eq.~(\ref{eq:expd}). Gaussian random numbers
[Eq.~(\ref{eq:gaussgauss})] can be produced with

\begin{verbatim}
 boost::normal_distribution<double> dist4(mu,sigma);
\end{verbatim}

\noindent where \texttt{mu} is the mean of the distribution and
\texttt{sigma} its width. Combining generators and distributions can
be accomplished with \texttt{boost::variate\_generator}. For example,
to produce 100 uniform random numbers in the interval $[0,1)$ using
the Mersenne Twister:

\SourceCodeLines{99}
\begin{Verbatim}[fontsize=\small]
 #include <boost/random.hpp>

 int main (void)
 {

 // define distribution
 boost::uniform_real<double> dist(0.,1.);

 // define the PRNG engine
 boost::mt19937 engine; 

 // create a normally-distributed generator
 boost::variate_generator<boost::mt19937&,
     boost::normal_distribution<double> >
     rng(engine,dist);
  
  // seed it
  engine.seed(1234u);

 // use it
 for (int i = 0; i < 100; i++){
     std::cout << rng() << "\n";
 }
\end{Verbatim}

\noindent For further details consult the Boost documentation
\cite{comment:boost}.

\subsection{GSL}

The GSL is similar to the Boost libraries. One can define both PRNG
engines and distributions and combine these to produce pretty much
any kind of random number.  For example, to produce 100 uniform random
numbers in the interval $[0,1)$ using the Mersenne Twister:

\SourceCodeLines{99}
\begin{Verbatim}[fontsize=\small]
 #include <stdio.h>
 #include <gsl/gsl_rng.h>

 int main()
 {
     gsl_rng *rng;                             /* pointer to RNG */
     int    i;                                       /* iterator */
     double u;                                  /* random number */

     rng = gsl_rng_alloc(gsl_rng_mt19937); /* allocate generator */
     gsl_rng_set(rng,1234)                 /* seed the generator */

     for(i = 0; i < 100; i++){
         u = gsl_rng_uniform(rng);    /* generate random numbers */
         printf("%f\n", u);
     }

     gsl_rng_free(rng);                      /* delete generator */

     return(0);
 }
\end{Verbatim}

\noindent For further details, check the GSL documentation
\cite{comment:gsl}.

\section{Random Numbers and cluster computing}

When performing simulations on large (parallel) computer clusters,
it is very easy to quickly use vast amounts of pseudo random
numbers. While some PRNGs are easily parallelized, others cannot be
parallelized at all. Some generators lose their efficiency and/or
the quality of the random numbers suffers when parallelized. It
goes beyond the scope of these lecture notes to cover this problem
in detail, however, the reader is referred to a detailed description
of the problems and their solution in Ref.\cite{mertens:09}.

The simplest (however not very rigorous) parallelization technique
is to have each process use the same PRNG, however with a different
seed. If the period of the PRNG is very large, one can hope to generate
streams of random numbers that do not overlap. In such a case, one can
either use a ``seed file'' where accounting of the used seeds
is done, or generate the seeds randomly for each process. A better
approach is either block splitting or leapfrogging where one random
number stream is used and distributed to all processes in blocks (block
splitting) or in a round-robin manner (leapfrogging) \cite{mertens:09}.

\subsection{Seeding}

In the case where the simulations are embarrassingly parallel
(independent simulations on each processor that do not communicate)
one has to be careful when choosing seeds on multi-core nodes. It
is customary to use the CPU time since January 1, 1970 with the
\texttt{time( )} command.  However, when multiple instances are started
on one node with multiple processor cores, all these will have the
{\em same} seeds because the \texttt{time( )} function call happens
for all jobs at once.  A simple solution is to combine the system
time with a number that is unique on a given node: the process ID
(PID). Below is an excerpt of a routine that combines the seconds since
January 1, 1970 with the PID using a small randomizer. Empirically,
there might be one seed collision every $10^4$ job submissions.

\SourceCodeLines{99}
\begin{Verbatim}[fontsize=\small]
 long seedgen(void)
 {
     long s, seed, pid;
 
     pid = getpid();        /* get process ID */
     s = time ( &seconds ); /* get CPU seconds since 01/01/1970 */

     seed = abs(((s*181)*((pid-83)*359))%104729);
     return seed;
 }
\end{Verbatim}

\section{Final recommendations}

Dealing with random numbers can be a delicate issue. Therefore \ldots

\begin{itemize}

\item[$\Box$]{Always try to run your simulations with two
different PRNGs from different families, at least for small testing
instances. One option would be to use an excellent but slow PRNG
versus a very good but fast PRNG. For the production runs then switch
to the fast one.}

\item[$\Box$]{To ensure data provenance always store the information
of the PRNG as well as the seed used (better even the whole code)
with the data. This will allow others to reproduce your results.}

\item[$\Box$]{Use trusted PRNG implementations. As much as it might
feel good to make your own PRNG, rely on those who are experts in
creating these delicate generators.}

\item[$\Box$]{Know your PRNG's limits: How long is the period? Are
there known problems for certain applications? Are there correlations
at any time during the sequence? \ldots}

\item[$\Box$]{Be careful with parallel simulations.}

\end{itemize}

\section*{Acknowledgments}

I would like to thank Juan Carlos Andresen, Ruben Andrist and Creighton
K.~Thomas for critically reading the manuscript.

\bibliographystyle{plain}  
\bibliography{comments,refs}

\begin{thebibliography}{10}
\bibitem{comment:period}
The period of a generator is the smallest integer $\rho>0$ such that the
  sequence of random numbers repeats after every $\rho$ numbers.
\end{thebibliography}

\end{document}